# Analysis of perturbed $H_2O$ vibrations beyond Fourier transform

*Walter Langel,*

Institut für Biochemie, Universität Greifswald, 17489 Greifswald, Germany

langel@uni-greifswald.de

New analysis methods for the vibrational dynamics from molecular dynamics are proposed and applied to liquid $H_2O$. The internal modes of $H_2O$ are amplitude-modulated by Langevin dynamics with the frequency of the $H_2O$ libration (~600 cm$^{-1}$) and its first overtone (~1200 cm$^{-1}$). The carrier signal is frequency-modulated by the variation of the hydrogen bond strength due to thermal motion and by dephasing collisions.

The standard power spectra of the bond parameters yield broadened normal modes due to a superposition of several perturbations. Here, the oscillating bond lengths and angles are demodulated by the spline interpolation of the minima or maxima. Additionally, a zero-crossing method is described, which yields the spectrum of the frequency modulated carrier without Fourier transform by directly converting the oscillation periods.

# I. INTRODUCTION

The high frequency internal vibrations of the O-H bond and the H-O-H bending of free $H_2O$ molecules are known from textbooks [1], but the spectra in liquids are strongly perturbed. Understanding the full vibrational spectrum of water is of great interest for many properties, including the anomalously high specific heat of about 9 R (gas constant), which is still subject to study [2],[3]. In the liquid the $H_2O$ molecules are trapped in cages and see a hindrance potential, which transfers the free rotations into librations around 600 cm$^{-1}$ [4]. Translational motions give rise to hydrogen network vibrations around 200 cm$^{-1}$.

Here, three perturbations that affect the spectrum are considered. The discrete molecules in the liquid show Brownian motion. The resulting viscous damping and the collisions with neighboring molecules are usually addressed as Langevin dynamics[5] which result in amplitude variations and the dephasing of the internal vibrations. In protic liquids, the hydrogen bonding of the H-atom with adjacent acceptors weakens the covalent O-H bond. The centers of the vibrational motions are oscillating due to thermal motion [6]. This oscillation modulates the strength of the hydrogen bonds of the respective H-atoms. The center motion is thus directly correlated to the strengths of the H-bond and of the covalent O-H-bond. In turn, the latter determines the vibrational frequency, resulting in the frequency modulation of the normal modes by hydrogen bond fluctuations. This so-called adiabatic coupling induces the frequency modulation (FM) of the O-H stretching vibration and the well-known broadening and redshift [7].

The trajectories from molecular dynamics simulations (MD) yield the time dependence of molecular parameters such as the bond lengths or angles. Many calculations on water with simple force fields employ rigid O-H bonds and cannot reproduce the influence of the hydrogen bonding on them. Force fields that are more sophisticated have to be parametrized against electron structure calculations. Static calculations, in turn, do not reflect the liquid dynamics. It is thus more straightforward to use the first principles molecular dynamics for tracing the influence of the hydrogen bonds on the covalent O-H bond lengths and on the vibrational spectra[8]. The power spectral densities (PSD) from these data has peaks at the normal modes [9]. The calculation of the PSD by Fourier analysis (fast Fourier transform, FFT) in general does not provide the possibility to distinguish between different perturbations, which result in broadening.

In communication technologies, the separation of the low and high frequency signals is well known. The modulation of the amplitude (AM) of a high frequency carrier by a low frequency signal results in side bands and thus in spectral broadening [10,11]. (This should not be confused with the beat phenomenon, where two oscillations with very similar frequencies are superposed resulting in a single carrier modulated with the small difference frequency). FM means that the carrier frequency varies according to the amplitude of the transmitted signal[11,12]. The AM and FM broadcasting is based on the unambiguous decomposition of the modulated signals into high frequency (HF) carrier and low frequency (LF) spectrum by various demodulation techniques.

Inspired by this, the perturbations of the spectra from first principles molecular dynamics of $H_2O$ in the liquid phase are analyzed here in a wide frequency range around and below the resonant transitions in the low frequency spectra. Beyond the simple Fourier transform, the PSD is obtained by three other methods, and it is demonstrated that additional information can be deduced.

## II. METHOD

In this study, 33 water molecules in the liquid phase in a cubic box with 1 nm side lengths were subjected to a first principles simulation using the Car-Parrinello method[13,14], which combines the description of the electron structure by the density functional theory (DFT) with classical nuclei dynamics. The Goedecker-Teter-Hutter pseudopotentials[15-17] and the Becke-Lee-Young-Parr functional (BLYP) for the exchange and correlation energies [18-20] were used, and the program sets up a plane wave basis, which had a cutoff of 60 Ryd in the present calculation. Some details of my simulations were described earlier [21,22]. The time step is 0.085 fs (3.5 a.t.u.), and the simulation extended to $5 \cdot 10^5$ steps or $t_{sim} = 42.5\, ps$ after equilibration. It is common practice to reduce the size of the trajectory by saving only a fraction of the calculated frames. Storing each tenth frame resulted in a sampling interval of dt = 0.85 fs here, which is still considerably shorter than in typical classical molecular dynamics runs, where the step width typically is 1 fs and only one out of 100 or 1000 frames is stored. The present dt corresponds to a Nyquist frequency of $\tilde{\nu}_{ny} = \frac{1}{2 \cdot dt \cdot c} = 2 \cdot 10^4\, cm^{-1}$, where c is the speed of light. This limit is far beyond the range of the IR-spectra. The molecular graphics program vmd [23] was used for extracting the bond lengths and angles from the trajectories as a function of time. Data were processed by a scilab script [24] and plotted with xmgrace

The power spectrum $P(\omega)$ of a motion along a coordinate r is given as a function of the angular frequency $\omega = 2 \cdot \pi \cdot c \cdot \tilde{\nu}$ from $P(\omega) = m \cdot \omega^2 \cdot \int \langle r(\tau) \cdot r(t+\tau) \rangle_\tau \exp(-i\omega t) dt$ [25]. This autocorrelation function is conveniently evaluated as the squared modulus of the fast Fourier transform (FFT) of r, $P(\omega) = m \cdot \omega^2 \cdot \left| \int r(t) \cdot \exp(-i\omega t) dt \right|^2$. The factor $\omega^2$ accounts for the increase in the energy content with the frequency of a vibration having a given amplitude. The power spectra of the O-H stretches and bending was averaged over all molecules. The same evaluation was carried out for 16 molecules only, which resulted only in a slight increase of the noise, showing that the sample was large enough. The length of the trajectory corresponds to 760 and 3800 oscillation periods for modes at 600 and 3000 cm$^{-1}$, respectively. The statistics of the data was thus considered to be sufficient for the evaluations presented here. The broadening of the Fourier transform spectra due to a finite $t_{sim}$ is negligible, with the lower limit of the line width being $\Delta\tilde{\nu} = \frac{\tilde{\nu}}{n} = \frac{\tilde{\nu} \cdot t_{osc}}{t_{sim}} = \frac{1}{c \cdot t_{sim}} = 0.8\, cm^{-1}$. There, n and $t_{osc}$ are the number and duration of the oscillation periods at the frequency $\tilde{\nu}$, respectively. The velocity autocorrelation function is obtained from the inverse FFT of $P(\omega)$,

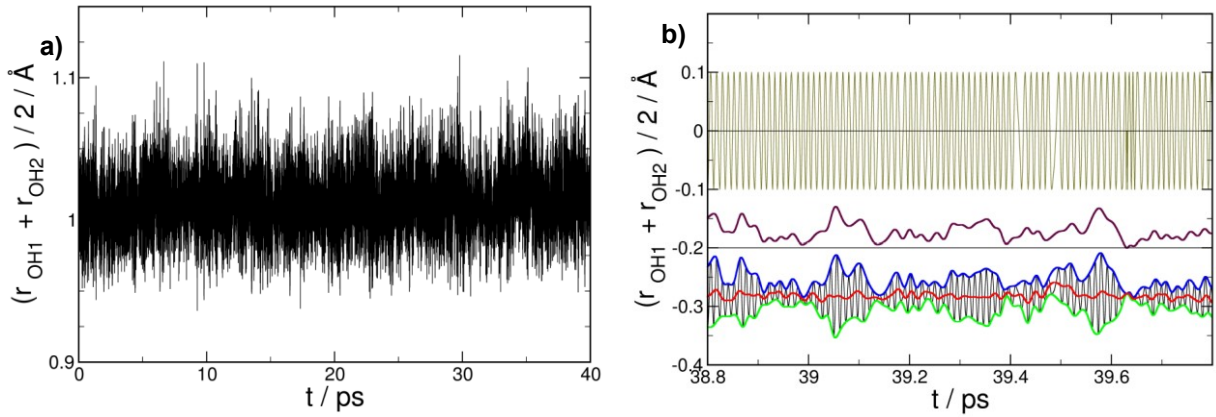

**Fig 1** Demodulation procedure:

**a**: The averaged O-H bond length in one $H_2O$-molecule f(t) as a function of time at 300±30 K over a 40 ps simulation time. Averaging over the two bonds enhances the contribution of the symmetric stretch mode to the signal.

**b:** A portion of the graph f(t) in **a** is zoomed in to a time span of 1 ps, and the demodulation procedure is visualized.

*From bottom to top:* (*black*) Data from **a**. The fluctuation of the minima and maxima of the oscillation cycles are clearly seen. The period of the oscillation roughly corresponds to the respective mode $\tilde{\nu}_{HF}$. The data are shifted by -1.3 Å for clarity. Spline interpolations $f_{max}(t)$ of the maxima (*blue*) and $f_{min}(t)$ of minima (*green*) are indicated. The center of the oscillation l(t) (*red*) fluctuates by approximately 0.03 Å around an average of 1.017 Å. This indicates elongation of the covalent bond through hydrogen bonding of the two H-atoms to other molecules with varying strength.

A separate plot of the oscillation amplitude a(t) (*maroon*) of f(t) above the bottom line (*black*) shifted to -0.2 Å . The amplitude varies between 0 and 0.1 Å.

The correction of f(t) for the fluctuations of the center (l(t)) and the amplitude (a(t)) results in the carrier oscillation only $f_{carrier}(t)$ (*olive*) with constant center 0 and amplitude 1, but with the same frequency as f(t). The time values $t_i$ for crossing the zero line (*black*), $f_{carrier}(t_i)=0$, now are well defined. Frequency variations and dephasing are clearly seen as irregularities in the oscillation.

## III. RESULTS

### A. Demodulation

**Fig 1a** shows the time dependence of the average O-H bond lengths. Instead of evaluating the individual bond lengths, the sums and differences of the two O-H bond lengths of each water molecule were taken, separating the symmetric and asymmetric stretching modes. The oscillation amplitude fluctuated in a wide range, but it is clearly seen that the center value also oscillated.

For obtaining signals without the carrier contributions, the minima and maxima, respectively, of each oscillation period were connected by two cubic spline interpolations on the time scale of data sampling, $f_{max}(t)$ and $f_{min}(t)$ (cf. **Fig 1**). The amplitudes, a(t), and the bond lengths, l(t), are obtained at each step as $a(t) = \frac{f_{max}(t) - f_{min}(t)}{2}$; $l(t) = \frac{f_{max}(t) + f_{min}(t)}{2}$. The function a(t) is then analogous to the modulating amplitude; in contrast, l(t) has no technical analog but reflects the fluctuation in the strengths of the hydrogen bonds. Separate LF power spectra are evaluated from a(t) and l(t). Obviously, the sampling interval for both is one vibrational period of the carrier signal, and the corresponding Nyquist frequency $\tilde{\nu}_{nyLF}$ is only half the carrier frequency $\tilde{\nu}_{HF}$, $\tilde{\nu}_{nyLF} = \frac{1}{2 \cdot dt_{HF} \cdot c} = \frac{\tilde{\nu}_{HF}}{2}$. For the data from the bending of the water molecules we have $\tilde{\nu}_{HF} \approx 1600\,cm^{-1}$ and $\tilde{\nu}_{nyLF} \approx 800\,cm^{-1}$, and for stretch, $\tilde{\nu}_{HF} \approx 3000\,cm^{-1}$ and $\tilde{\nu}_{nyLF} \approx 1500\,cm^{-1}$.

The carrier is $f_{carrier}(t) = \frac{f(t)-l(t)}{a(t)}$. This is analogous to filtering the AM in telecommunications; however, here, the carrier is still frequency modulated, and its frequency varies significantly (**Fig 1b**). The respective Fourier transform is addressed as the HF part of the spectrum. These operations are linear and should not give rise to overtones.

The O-H stretch oscillations around 3000 cm$^{-1}$ are not exactly reproduced when sampling in discrete intervals, and the calculated maxima and minima of the periodic function may thus deviate from the theoretical values. For a test sinus function, $f_s(t) = 1 A° \cdot \sin(2 \cdot \pi \cdot 3000\, cm^{-1} \cdot c \cdot t)$, a(t) and l(t) have constant theoretical values of 1 Å and 0, respectively. After sampling with dt=0.85 fs, one obtains the averages of $\bar{l} = -5 \cdot 10^{-6} \pm 8.3 \cdot 10^{-3} A°$ and $\bar{a} = 0.9903 \pm 2 \cdot 10^{-3} A°$ over 40 ps. This scatter may be reduced by applying cubic spline interpolation of f$_s$(t) on time steps of 0.1 fs and taking the minima and maxima of the oscillation from the interpolated curve. The spline resulted in a smoother approximation to the theoretical sine function and improved the numeric results for center and amplitude of the sine wave to $\bar{l} = -4 \cdot 10^{-8} \pm 0.12 \cdot 10^{-3} A°$ and $\bar{a} = 0.9997 \pm 0.027 \cdot 10^{-3} A°$. As a trial, this procedure was also applied to the water trajectory, and minima and maxima of each oscillation period for calculating f$_{min}$(t) and f$_{max}$(t) were taken from the spline rather than from the original data, respectively. This was a check, if the typical sampling interval of 0.85 fs resulted in spurious contributions to the spectra of the H$_2$O.

An alternative approach to the HF spectrum is independent from FFT and aims at separating the effects of the adiabatic coupling of the hydrogen bond from the dephasing of the oscillation. The time differences $\Delta t_i = t_{i+1} - t_i$ between two consecutive zero crossings t$_{i+1}$ and t$_i$ in the corrected signal (**Fig 1**) are converted to frequencies using $\tilde{\nu}_i = \frac{1}{2 \cdot \Delta t_i \cdot c}$. From n≈1900 (1600 cm$^{-1}$) or 3500 (3000 cm$^{-1}$) cycles during t$_{sim}$, 2n-1 frequencies are obtained, from which distributions with a binning of 10, 20, and 40 cm$^{-1}$ were calculated and referred to as the zero crossing spectra. As this operation may be non linear, its results have to be validated by comparing with FFT.

B. Bond parameters

The probability distributions of the O-H bond elongation x are well described by a Gaussian with a width corresponding to the temperature T, $L(x) = \exp\left(-\frac{1}{2} \cdot \frac{\mu_{OH} \cdot \omega^2}{R \cdot T} \cdot x^2\right)$, where the reduced mass $\mu_{OH}$ of the oscillator is set equal to the proton mass. L(x) is a Boltzmann distribution of the potential energy only. An unperturbed classic oscillator with the conservation of the sum of kinetic and potential energies has a distribution L`(x) of the elongation with the maxima at the turning points. The perturbations mentioned above, damping, dephasing and hydrogen bond fluctuations contribute to spreading L(x) out to a Gaussian. One may interpret this as a decoupling of the kinetic and potential energies. The velocities of the O and H atoms vary randomly due to collisions with ambient water molecules. This results in independent distributions of the kinetic and potential energies of the oscillators.

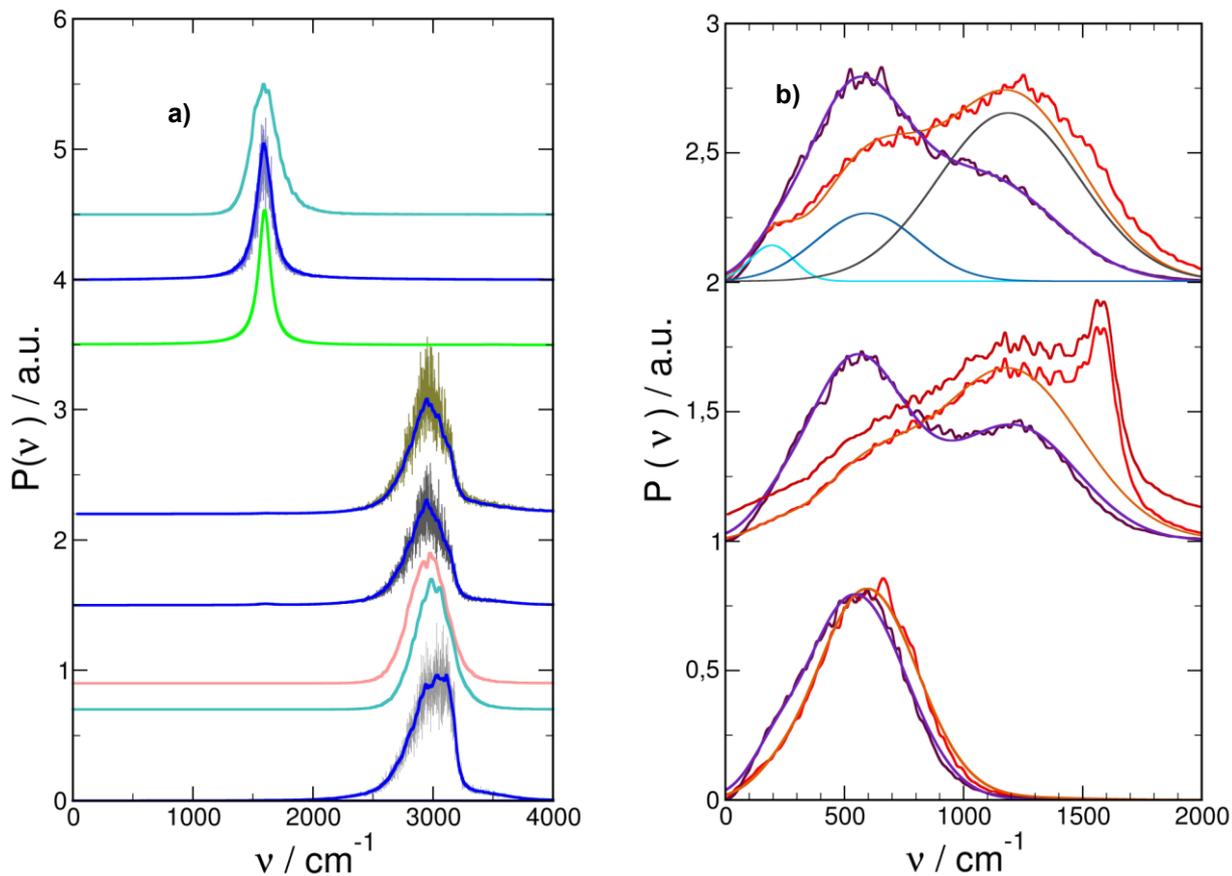

**Fig 2:** Spectra of liquid $H_2O$:

**a:** Peaks are presented in the range of the $H_2O$ normal modes. Traces are shifted in the y-direction and rescaled for clarity.

*From bottom to top:* (*gray*) A standard Fourier transform spectrum of the differences in the O-H bond lengths emphasizing the asymmetric stretching mode. The noise is strongly reduced after smoothing by convolution with a Gaussian with a 20 cm$^{-1}$ fwhm (*blue*).

HF spectra from zero crossing of the asymmetric (*cyan*) and symmetric (*pink*) stretch modes with bins of 20 cm$^{-1}$. This evaluation is completely independent of the Fourier transform, but yields a contiguous intensity in the same range with a similar overall shape. The zero crossing spectra reproduce the slight redshift of the symmetric vs. asymmetric mode.

The (*gray*, *blue*) as the bottom trace show the average of the O-H lengths yielding mainly symmetric stretch mode. The spectra for symmetric and asymmetric stretching overlap, but it is obvious from the smoothed data that the symmetric stretching is slightly red-shifted with respect to the asymmetric one, as expected.

The (*olive*, *blue*) traces are shown as before, using carrier only signal rather than the full signal. Although the amplitude modulation is suppressed here, no significant narrowing of the O-H vibration is observed with respect to the amplitude modulated signal (s. text).

The (*green*) trace describes the H-O-H bending mode by a damped oscillation $I(\tilde{\nu})$ with $\omega_0$=1600 cm$^{-1}$ and $\lambda$=0.074 (s. text).

The (*gray, violet*) is the Fourier spectrum of the H-O-H angles. Because only one normal mode (bending) contributes to this peak, it is highly symmetric and narrower than those from the two stretching modes.

The top trace (*cyan*) is the zero crossing spectrum of the bending mode. The position and shape of this mode are well reproduced, but the feature is slightly broader than the smoothed FFT curve.

**b:** A linear plot of low frequency spectra for (*red*) center and (*maroon*) amplitude oscillations smoothed by Gaussians with 20 cm$^{-1}$ fwhm is shown. All data are fitted by sums (*orange, dark blue*) of three Gaussians (s. text). *From bottom to top*: bending, symmetric, and asymmetric stretch.

(*middle*) The method of spline interpolation of raw data with dt=0.1 fs for smoothing time dependent curves and reducing scatter of minima and maxima was applied to center of symmetric mode before calculating LF spectra (*dark red*) (s. text). As the difference is not significant, all other data were evaluated with 0.85 fs sampling only (*red*).

(*top*) The three Gaussians as used for the fit are indicated for the center oscillation of the asymmetric stretch as H-bond network vibrations (*cyan*), libration (*blue*) and libration overtone (*gray*). The fwhm of the Gaussians are 215 cm$^{-1}$, 498 cm$^{-1}$ and 704cm$^{-1}$ for all hydrogen bonding vibrations, librations and overtones, respectively. The contribution of the feature close to 200 cm$^{-1}$ is most obvious for the center oscillations of the asymmetric mode.

The table quotes frequencies and intensities normalized to the respective librations.

| Assignment | Frequency of amplitude oscillation (*maroon*) | Intensity relative to libration | Frequency of center oscillation (*red*) | Intensity relative to libration |
|---|---|---|---|---|
| Perturbation of the asymmetric stretch mode | | | | |
| Hydrogen bond | 195 | 0.029 | 195 | 0.14 |
| Libration | 540 | 1 | 595 | 1 |
| Overtone | 1080 | 0.80 | 1190 | 2.4 |
| Perturbation of the symmetric stretch mode | | | | |
| Hydrogen bond | 195 | 0.025 | 195 | 0.059 |
| Libration | 546 | 1 | 595 | 1 |
| Overtone | 1092 | 0.90 | 1190 | 3.7 |
| Perturbation of the bending mode | | | | |
| Hydrogen bond | 195 | 0.045 | 195 | 0.016 |
| Libration | 545 | 1 | 595 | 1 |
| Overtone | -- | 0 | 1190 | 0.017 |

## C. Broadening of the normal modes

The bond lengths and angle fluctuations yield the stretching (~3000 cm$^{-1}$), and the bending (~1600 cm$^{-1}$) modes (**Fig 2a**). The linear plots of all of the spectra reveal significant scatter, and the smoothing by convolution with a Gaussian of 20 cm$^{-1}$ full width at half maximum (fwhm) suppresses the noise very efficiently without significant peak broadening. In the range of the internal H$_2$O modes, carrier-only spectra (cf. **Fig 1b**) are very similar in shape and signal-to-noise ratio to those of the full signal, and smoothing again is efficient (**Fig 2a**). The zero crossing spectra of the stretching modes were calculated without any reference to FFT, but still coincide very well with the smoothed HF-spectra.

## D. Low frequency spectra

Linear plots of the center and amplitude spectra reveal additional broad transitions below 1600 cm$^{-1}$ (**Fig 2b**). They were not identified in the spectra of the full data set before the extraction of LF because the intensity in the respective frequency ranges was too weak. The spectra are fitted by sums of three Gaussians with different weights (cf. **Fig 2b,** *top*). The fits have high correlation coefficient around 0.998-0.999, apart of the two fits of the center spectra of the stretch modes (*middle* and *top, red*). Here, the three peaks do not approximate additional intensities at higher frequencies. The center spectrum of the symmetric stretch contains a sharp edge at 1800 cm$^{-1}$ beyond the respective Nyquist frequency, indicating truncation of a feature with an even higher frequency.

The spectra from the fluctuation of the amplitude of H-O-H angle and of its center contain broad intensities centered at 550 cm$^{-1}$ and 600 cm$^{-1}$, respectively, and nearly coincide apart of a 50 cm$^{-1}$ shift The trailing edges of the features smoothly decrease to zero at about 1100 cm$^{-1}$.

In each of the two LF stretching spectra a further broad transition appears at approximately 1100-1200 cm$^{-1}$. It is fitted by a Gaussian with twice the center frequency and the squared variance as the respective one at 550-600 cm$^{-1}$ (**Fig 2b**). As the Gaussian at the higher frequency is the self-convolution of the lower one, this feature around 1200cm$^{-1}$ is considered as its overtone. These features have similar intensities as the respective fundamentals and exceed them even in the center spectra (**Fig 2b**). A small intensity around 200 cm$^{-1}$ merges with the leading tail of 600 cm$^{-1}$ feature. It cannot be described by the Gaussians at 550-600 cm$^{-1}$, and is reproduced by a third one at 195 cm$^{-1}$ (**Fig 2b**).

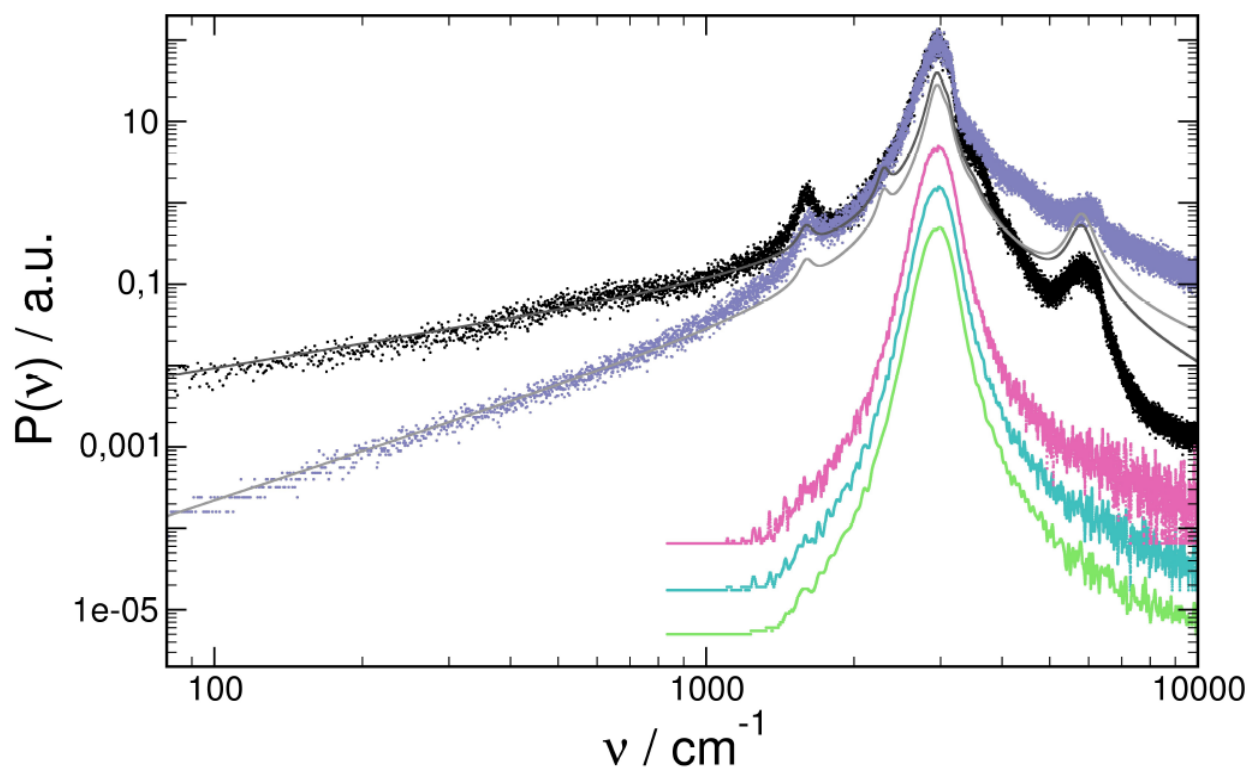

**Fig 3:** Double logarithmic plots of the power spectra of the average O-H bond length over the frequency without smoothing.

*From bottom to top*:

Zero crossing spectra with bins of 10 (*green*), 20 (*blue*) and 40 cm$^{-1}$ (*pink*). The peaks are slightly broader than in FFT, but the line shapes are identical, and no obvious increase of the width with the increasing binning interval is seen. All curves have a high signal to noise ratio, which is still improved in the very weak tails by increasing the binning interval.

The full spectrum (*black, dots*) from FFT. Intensities below 1000 cm$^{-1}$ amount less than 10$^{-3}$ of the peak and are only visible in the logarithmic plot. The suppression of this weak LF spectrum after the demodulation is clearly seen in the HF carrier spectrum (*gray, dots*), but the normal modes are very similar in both spectra. The full lines are guides to the eye and indicate $P_{HF}(\tilde{v}) = \tilde{v}^2 \cdot I(\tilde{v})$ (*black*) and $P_{full}(\tilde{v}) = \tilde{v} \cdot I(\tilde{v})$ (*gray*), s. text. The parameters for the maxima of $I(\tilde{v})$ (s. text) are:

| $\tilde{v}$ / $cm^{-1}$ | Assignment | I | λ |
|---|---|---|---|
| 1600 | Bending mode | 0.002 | 0.074 |
| 2300 | Side band | 0.02 | 0.06 |
| 2950 | Symmetric stretching mode | 1.0 | 0.08 |
| 3100 | Asymmetric stretching mode | 0.35 | 0.06 |
| 3500 | Side band | 0.01 | 0.04 |
| 5800 | Overtone of stretching modes | 0.01 | 0.12 |

The double logarithmic plot shows the PSD in large frequency and intensity ranges beyond the resonant transitions (**Fig 3**). The spectrum of the sum of the O-H elongations contains a strong feature near 3000 cm$^{-1}$ for the stretching modes, which is fitted here by two broad peaks at 2950 and 3100 cm$^{-1}$. These calculated frequencies diverge from experimental data showing the O-H stretching modes between 3200 and 3400 cm$^{-1}$[26], possibly indicating overestimation of the hydrogen bonding strength by the DFT methods applied here.

Reproducing the spectrum using damped harmonic modes (s. discussion) affords the additional weak maxima, which are not seen in the linear plots. The edges of the peak near 3000 cm$^{-1}$ were tentatively approximated by two weak features at 2300 and 3500 cm$^{-1}$, which are assigned to sidebands of the stretching peak due to AM by the librations around 600 cm$^{-1}$ (**Fig 2b** (*middle*)). A peak at 1600 cm$^{-1}$ is assigned to the bending mode. Due to the molecular symmetry, the bending mode mixes with the symmetric stretching spectrum as shown here but not with the asymmetric stretching. Furthermore, the logarithmic spectrum contains a small maximum at 5800 cm$^{-1}$, which is assigned to an overtone of the stretching modes.

The power spectra are related to velocity autocorrelation functions, which show a very fast decay within approximately 0.1 ps and a fluctuating, very slowly decaying periodic function.

## IV. Discussion

In this paper, new methods for the separation of the high and low frequency signals from molecular dynamics trajectories are proposed. Five types of spectra are compared: the conventional power spectrum of the signal, the spectrum resulting from filtering out the amplitude variation, the zero crossing spectrum and two low frequency spectra, which are derived from the variations of the oscillation amplitude and centers positions, respectively, of the bond vibrations. The influence of the external modes of the molecule in its cage (librations, hydrogen bond network motions) and of the cage fluctuation (Langevin dynamics with viscous damping and dephasing by collisions) are now discussed.

E. Viscous damping

In the spectra of the full signal (**Figs 2a, 3**), the typical infrared transitions are broadened. Inhomogeneous contributions may be discarded in the liquid and the simplest approximation is given by a sum of the damped oscillations, $I(\tilde{v}) = \sum_i \frac{I_i(\tilde{v}_i)}{(\tilde{v}^2 - \tilde{v}_i^2)^2 + \lambda^2 \cdot \tilde{v}_i^2 \cdot \tilde{v}^2}$. The center $\tilde{v}_i$ is a resonant vibrational transition of the molecule, and the dimensionless damping parameter λ determines the width of the mode. Ignoring broadening resulting from the hydrogen bond and dephasing, λ is only related to the mobility, µ, of the moving particle, and by Stoke's law, to the viscosity η of the liquid:

$$\lambda \cdot \omega_i = \lambda \cdot 2 \cdot \pi \cdot c \cdot \tilde{v}_i = \frac{1}{\mu \cdot m} = \frac{6 \cdot \pi \cdot \eta \cdot r}{m} \Rightarrow \eta = \frac{m \cdot \lambda \cdot 2 \cdot \pi \cdot c \cdot \tilde{v}_i}{6 \cdot \pi \cdot r}$$

With the molar mass m=18 g/mol and the radius r=0.19 nm for a nearly spherical water molecule, the observed broadening with $\lambda = 0.07 - 0.08$ affords viscosities of $\eta = 3.3 - 6.1 \cdot 10^{-3}$ Pa·s. They are significantly larger than the results on water from experiment (0.9·10⁻³ Pa·s) and from similar first principles MD, as in the present work (2.1·10⁻³ Pa·s) [27]. Obviously, viscous damping cannot fully account for the observed width; therefore, hydrogen bonding and dephasing are assumed to broaden the transitions significantly.

A damped oscillator spectrum would be nearly constant below 1000 cm⁻¹. The full spectrum is very weak in this range but increases with $P_{full}(\tilde{v}) \propto \tilde{v}$ (**Fig 3**). The HF spectrum is even weaker, but increases with $P_{HF}(\tilde{v}) \propto \tilde{v}^2$. Consequently, the functions $P_{HF}(\tilde{v}) = \tilde{v}^2 \cdot I(\tilde{v})$ and $P_{full}(\tilde{v}) = \tilde{v} \cdot I(\tilde{v})$ are better approximations than $I(\tilde{v})$ to the respective data in an extended frequency range. The damped oscillator model makes a very simple assumption on the velocity autocorrelation function of the vibrating system, providing a periodic function multiplied with a single exponential. The complicated liquid dynamics obviously results in a deviation from this simple assumption and generates a modified spectrum. Due to the resonant modes in the water molecule, $P(\tilde{\nu})$ is not either consistent with typical Brownian noise spectra[28], where the intensity decreases with increasing frequency.

### F. Amplitude modulation

Amplitude fluctuations from interactions with surrounding particles result in the LF amplitude spectrum of the vibrations (**Fig 2b**). Using only the carrier signal suppresses the eventual effect of the AM to the normal mode spectrum, especially to the side bands (**Fig 2a**). The similarity of this spectrum to the full signal shows that amplitude modulation does not significantly perturb the observed vibrational transitions.

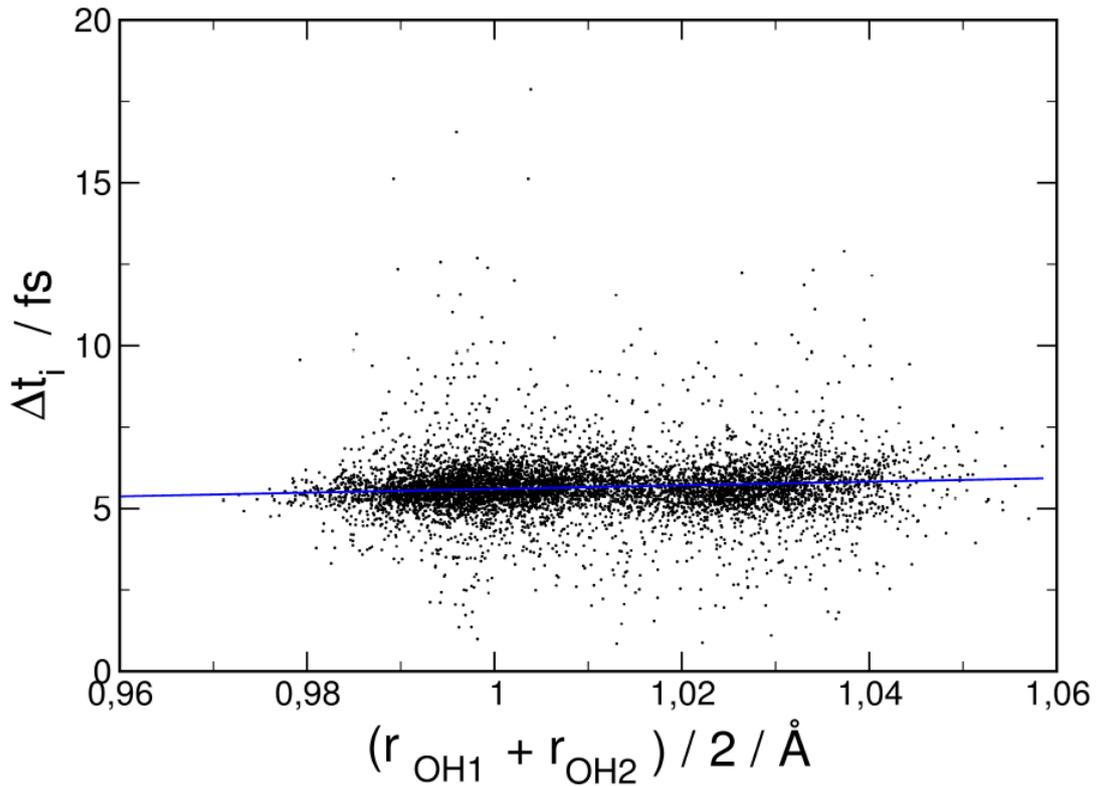

**Fig 4:** The distance between the zero crossings in the HF signal of symmetric stretching vs. O-H bond lengths. The (black) data points show large scatter, (see text) and the positive slope of the linear regression (blue) through the data points indicates some correlation between the vibrational period and the bond length.

### G. Hydrogen bonding

By adiabatic coupling, the power spectra of the normal modes are broadened. The carrier time function $f_{carrier}(t)$ reflects this as a variation of the oscillation time (**Fig 1b**). As the broadening of the normal modes by AM is small, FM resulting from hydrogen bonding probably is the main contribution to the width of the normal modes. The low frequency spectra of the oscillations of the vibrational center show that the strength of the hydrogen bond varies as a result of the thermal motion of the adjacent molecules. From the zero crossing data, the influence of the hydrogen bonding on the carrier frequency is directly traced by plotting the time between two zeros over the actual bond length (**Fig 4**). This shows an obvious correlation, and with increasing bond length, the time is increasing. By increasing the hydrogen bond strength, the length of the covalent O-H bond increases and the stretching frequency decreases. This redshift means an increase in the vibrational period.

### H. Dephasing

The carrier oscillation is affected by dephasing collisions. In the time dependence of the elongation, the respective phase jumps are observed (**Fig 1b**). This may clearly be distinguished from the damped oscillations with smoothly varying vibrational periods in the same traces. Dephasing of the vibrations by collisions with adjacent water molecules has a different influence on the two methods for converting the carrier into the HF spectrum.

By the FFT, the influences of damping and dephasing cannot distinguished, but both contribute to the broadening of the resonant modes. The zero crossing method gives one frequency value for each single half period of oscillation. For dephasing collisions, single frequency values are obtained, which are much further away from the unperturbed vibration; whereas the Fourier method tends to average over all oscillation periods. Zero crossing should be more sensitive to dephasing and this effect should yield widely broadened spectra. Indeed, the scatter in **Fig 4** may be attributed to the dephasing collisions. The frequency distribution from these data yields zero crossing spectra, which are slightly broader than the respective smoothed Fourier spectra (**Figs 2a, 3**), and this does not seem to be due to the binning widths.

## I. Assignment of low frequent modes

The strong peak around 550-600 cm$^{-1}$ is assigned to the perturbation of the normal modes of $H_2O$ by the librations of the molecules with respect to their cages. In the spectrum of the H-O-H angle (**Fig 2b**, bottom), this is the strongest feature, and in the other spectra it is also well defined. During experimental studies [29], and in earlier first principles MD [4] on water, the 600 cm$^{-1}$ mode was also found. These spectra directly reflect the dipole autocorrelation of the molecule and thus directly reproduce the libration frequency.

Here, the perturbations of the internal modes by the hindrance potential of the cage stopping free rotation are analyzed. Only if this potential has one-fold symmetry, the perturbation of the bond lengths has the same frequency as the libration itself. A potential with two-fold symmetry applies the same perturbation to the O-H bond length twice during each oscillation period of the libration and thus gives rise to a perturbation with the doubled frequency. The feature around 1200 cm$^{-1}$ does not seem to occur in the dipole autocorrelation spectra [4], which is consistent with its assignment as overtone. The hindrance potentials may thus result from a superposition of components with one- and two-fold symmetry, and the latter may be related to the symmetry of the $H_2O$-molecule.

The feature around 200 cm$^{-1}$ is assigned to the hydrogen bonding network modes of liquid $H_2O$ [4]. The relative intensities of libration and hydrogen network motions are possibly not related to the respective densities of states, since the oscillations are not observed directly, but only as perturbations of normal modes.

## V. CONCLUSION

First principles molecular dynamics straightforward reproduce the full spectra of molecular motion including hydrogen bonding. The dynamics in the liquid result from several contributions, perturbing these normal modes and generating LF center and amplitude spectra. As they are due to two different mechanisms, they do not fully converge:

- External rotations and translations of the $H_2O$-molecule convert into molecular librations and hydrogen bond vibrations, respectively, due to the hindrance potential in the cage. This also perturbs the center of mass motion of $H_2O$ by Langevin dynamics (viscous damping, collision dephasing) with frequencies of libration and first overtone. The amplitudes of bending and stretching vibrations are affected. This AM yields very weak side bands of the stretching mode and mainly as LF spectrum of amplitudes, but does not account for the well-known broadening of the water stretching modes by hydrogen bonding.

- The molecular motion in the cage further induces a periodic modulation of the hydrogen bond strength in protic systems. This has two effects: The covalent bond length is modulated, which is seen as center spectra, and the frequency of the stretch frequency shifts in phase with the variation of the hydrogen bond strength (adiabatic coupling). This is seen as FM of the internal modes and results in significant broadening.

The standard method for calculating spectra from molecular dynamics results applies the FFT of molecular parameters as a function of time. This way, the separation of different contributions to the spectra is not possible, but all perturbations mix and result in a broadening of the resonant transitions, and only very weak LF intensities. Here, methods beyond a direct calculation of the power spectra are described and applied to liquid water. This is a prototypical approach for the perturbation of vibrations by collisions and variations of the hydrogen bond strength, and the analysis should be transferable to systems that are more complex. Different contributions to the spectra are separated by the demodulation of the signal applying spline interpolation of the maxima or minima. Such spectra should help to clarify the complex dynamics of liquid water, which are possibly related to the high specific heat in the liquid. The second new algorithm is the frequency analysis of the carrier using zero crossing distances. This approach is fully independent of FFT, but yields similar results and is, therefore, validated here.

## VI. ACKNOWLEDGMENTS

The water trajectory was provided by Daniel Möller and Norman Geist (Greifswald). I further acknowledge useful discussions with Dr. Susan Köppen (Bremen).